\documentstyle[11pt,aaspp4]{article}

\slugcomment{Accepted by The Astronomical Journal}

\lefthead{Schneider et al.}
\righthead{HET Observations of L Dwarfs}

\begin{document}

\title{L Dwarfs Found in Sloan Digital Sky Survey Commissioning Data~II.
\hbox{Hobby-Eberly Telescope} Observations
\footnote{Based on observations obtained with the Sloan Digital
 Sky Survey, which is
 owned and operated by the Astrophysical Research Consortium, and
 based on observations obtained with the Hobby-Eberly
 Telescope, which is a joint project of the University of Texas at Austin,
 the Pennsylvania State University, Stanford University,
 Ludwig-Maximillians-Universit\"at M\"unchen, and Georg-August-Universit\"at
 G\"ottingen.}
}
\author{
Donald~P.~Schneider\altaffilmark{\ref{PennState}},
%Scott~F.~Anderson\altaffilmark{\ref{UW}},
Gillian~R.~Knapp\altaffilmark{\ref{Princeton}},
Suzanne~L.~Hawley\altaffilmark{\ref{UW}},
Kevin~R.~Covey\altaffilmark{\ref{UW}},
Xiaohui~Fan\altaffilmark{\ref{IAS}},
Lawrence~W.~Ramsey\altaffilmark{\ref{PennState}},
Gordon~T.~Richards\altaffilmark{\ref{PennState}},
Michael~A.~Strauss\altaffilmark{\ref{Princeton}},
%Daniel~E.~Vanden~Berk\altaffilmark{\ref{FNAL}},
%David~A.~Golimowski\altaffilmark{\ref{JHU}},
James~E.~Gunn\altaffilmark{\ref{Princeton}},
%David~Schlegel\altaffilmark{\ref{Princeton}},
%Wolfgang~Voges\altaffilmark{\ref{MPE}},
%Brian~Yanny\altaffilmark{\ref{FNAL}},
Gary~J.~Hill\altaffilmark{\ref{Texas}},
Phillip~J.~MacQueen\altaffilmark{\ref{Texas}},
Mark~T.~Adams\altaffilmark{\ref{Texas}},
%John~A.~Booth\altaffilmark{\ref{Texas}},
Grant~M.~Hill\altaffilmark{\ref{Texas}},
\v Zeljko~Ivezi\'c\altaffilmark{\ref{Princeton}},
Robert~H.~Lupton\altaffilmark{\ref{Princeton}},
Jeffrey~R.~Pier\altaffilmark{\ref{USNOAZ}},
David~H.~Saxe\altaffilmark{\ref{IAS}},
Matthew~Shetrone\altaffilmark{\ref{Texas}},
Joseph~R.~Tufts\altaffilmark{\ref{Texas}},
Marsha~J.~Wolf\altaffilmark{\ref{Texas}},
%John~E.~Anderson,~Jr.\altaffilmark{\ref{FNAL}},
%Neta~A.~Bahcall\altaffilmark{\ref{Princeton}},
%S.~Bastian\altaffilmark{\ref{FNAL}},
%R.~Bender\altaffilmark{\ref{Munich}},
%E.~Berman\altaffilmark{\ref{FNAL}},
J.~Brinkmann\altaffilmark{\ref{APO}},
%Robert~Brunner\altaffilmark{\ref{Caltech}},
Istv\'an~Csabai\altaffilmark{\ref{JHU}}$^,$\altaffilmark{\ref{Hungary}},
%Mamoru~Doi\altaffilmark{\ref{Tokyo2}},
%M.~Eracleous\altaffilmark{\ref{PennState}},
%G.~Federwitz\altaffilmark{\ref{FNAL}},
%Masataka~Fukugita\altaffilmark{\ref{CosJapan}}$^,$\altaffilmark{\ref{IAS}},
%V.~Gurbani\altaffilmark{\ref{FNAL}},
G.S.~Hennessy\altaffilmark{\ref{USNODC}},
%Robert~B.~Hindsley\altaffilmark{\ref{NRL}},
%Peter~J.~Kunst\altaffilmark{\ref{JHU}},
%Donald~Q.~Lamb\altaffilmark{\ref{Chicago}},\altaffilmark{\ref{Fermi}}
%C.~Lindenmeyer\altaffilmark{\ref{FNAL}},
%Jon~Loveday\altaffilmark{\ref{Sussex}},
%P.~Mantsch\altaffilmark{\ref{FNAL}},
%Timothy~A.~McKay\altaffilmark{\ref{Michigan}},
%Jeffrey~A.~Munn\altaffilmark{\ref{USNOAZ}},
%C.~Nance\altaffilmark{\ref{Texas}},
%T.~Nash\altaffilmark{\ref{FNAL}},
%S.~Okamura\altaffilmark{\ref{Tokyo}},
%R.C.~Nichol\altaffilmark{\ref{CMU}},
%R.~Rechenmacher\altaffilmark{\ref{FNAL}},
%B.~Rhoads\altaffilmark{\ref{Texas}},
%C.H.~Rivetta\altaffilmark{\ref{FNAL}},
%E.L.~Robinson\altaffilmark{\ref{Texas}},
%B.~Roman\altaffilmark{\ref{Texas}},
%G.~Sergey\altaffilmark{\ref{FNAL}},
%C.~Stoughton\altaffilmark{\ref{FNAL}},
%G.P.~Szokoly\altaffilmark{\ref{Potsdam}},
%Aniruddha~R.~Thakar\altaffilmark{\ref{JHU}},
%D.L.~Tucker\altaffilmark{\ref{FNAL}},
%G.~Wesley\altaffilmark{\ref{Texas}},
%J.~Willick\altaffilmark{\ref{Stanford}},
%P.~Worthington\altaffilmark{\ref{Texas}},
and
Donald~G.~York\altaffilmark{\ref{Chicago}}$^,$\altaffilmark{\ref{Fermi}}
}

%email addresses: dps@astro.psu.edu,
%slh@astro.washington.edu,
%gk@astro.princeton.edu,

% Notice that each of these authors has alternate affiliations, which
% are identified by the \altaffilmark after each name.  The actual alternate
% affiliation information is typeset in footnotes at the bottom of the
% first page, and the text itself is specified in \altaffiltext commands.
% There is a separate \altaffiltext for each alternate affiliation
% indicated above.

\newcounter{address}
\setcounter{address}{2}
\altaffiltext{\theaddress}{Department of Astronomy and Astrophysics, The
   Pennsylvania State University, University Park, PA 16802.
\label{PennState}}
\addtocounter{address}{1}
\altaffiltext{\theaddress}{Princeton University Observatory, Princeton,
   NJ 08544.
\label{Princeton}}
\addtocounter{address}{1}
\altaffiltext{\theaddress}{University of Washington, Department of
   Astronomy, Box 351580, Seattle, WA 98195.
\label{UW}}
\addtocounter{address}{1}
\altaffiltext{\theaddress}{The Institute for Advanced Study, Princeton,
   NJ 08540.
\label{IAS}}
%\addtocounter{address}{1}
%\altaffiltext{\theaddress}{Fermi National Accelerator Laboratory, P.O. Box 500,
%   Batavia, IL 60510.
%\label{FNAL}}
%\addtocounter{address}{1}
%\altaffiltext{\theaddress}{Max-Planck-Institute f\"ur extraterrestrische
%   Physik,
%   Geissenbachstr.~1, D-85741 Garching, Germany.
%\label{MPE}}
%\addtocounter{address}{1}
%\altaffiltext{\theaddress}{Astronomy Department, California Institute of
%   Technology, Pasadena, CA 91125.
%\label{Caltech}}
%\addtocounter{address}{1}
%\altaffiltext{\theaddress}{Department of Astronomy and Research Center for the
%   Early Universe, School of Science, University of Tokyo, Hongo, Bunkyo,
%   Tokyo 113-0033, Japan.
%\label{Tokyo}}
%\addtocounter{address}{1}
%\altaffiltext{\theaddress}{Department of Astronomy and Research Center for the
%   Early Universe, School of Science, University of Tokyo, Mitaka,
%   Tokyo 181-0015, Japan.
%\label{Tokyo2}}
\addtocounter{address}{1}
\altaffiltext{\theaddress}{Department of Astronomy, McDonald Observatory,
   University of Texas, Austin, TX~78712.
\label{Texas}}
\addtocounter{address}{1}
\altaffiltext{\theaddress}{US Naval Observatory, Flagstaff Station,
   P.O. Box 1149, Flagstaff, AZ 86002-1149.
\label{USNOAZ}}
%\addtocounter{address}{1}
%\altaffiltext{\theaddress}{Universit\"ats-Sternwarte,
%   Scheinerstrasse~1, 81679~M\"unchen, Germany.
%\label{Munich}}
%\addtocounter{address}{1}
%\altaffiltext{\theaddress}{Remote Sensing Division, Code~7215, Naval Research
%   Laboratory, 4555~Overlook~Ave. SW, Washington, DC~20375.
%\label{NRL}}
%\addtocounter{address}{1}
%\altaffiltext{\theaddress}{Institute for Cosmic Ray Research, University
%   of Tokyo, Midori, Tanashi, Tokyo 188-8588, Japan
%\label{CosJapan}}
\addtocounter{address}{1}
\altaffiltext{\theaddress}{Apache Point Observatory, P.O. Box 59,
   Sunspot, NM 88349-0059.
\label{APO}}
\addtocounter{address}{1}
\altaffiltext{\theaddress}{Department of Physics and Astronomy,
   Johns Hopkins University, 3701 University Drive, Baltimore, MD 21218.
\label{JHU}}
\addtocounter{address}{1}
\altaffiltext{\theaddress}{Department of Physics of Complex Systems,
   E\"otv\"os University, P\'azm\'ay P\'eter \hbox{s\'et\'any 1/A,}
   H-1117, Budapest, Hungary.
\label{Hungary}}
\addtocounter{address}{1}
\altaffiltext{\theaddress}{US Naval Observatory, 3450 Massachusetts Avenue NW,
   Washington, DC 20392-5420.
\label{USNODC}}
\addtocounter{address}{1}
\altaffiltext{\theaddress}{Astronomy and Astrophysics Center, University of
   Chicago, 5640 South Ellis Avenue, Chicago, IL 60637.
\label{Chicago}}
\addtocounter{address}{1}
\altaffiltext{\theaddress}{The University of Chicago, Enrico Fermi Institute,
  5640 South Ellis Avenue, Chicago, IL 60637.
\label{Fermi}}
%\addtocounter{address}{1}
%\altaffiltext{\theaddress}{Astronomy Centre, University of Sussex, Falmer,
%   Brighton BN1 9QJ, UK.
%\label{Sussex}}
%\addtocounter{address}{1}
%\altaffiltext{\theaddress}{Department of Physics,University of Michigan,
%   500 East University, Ann Arbor, MI 48109.
%\label{Michigan}}
%\addtocounter{address}{1}
%\altaffiltext{\theaddress}{Dept. of Physics, Carnegie Mellon University,
%     5000~Forbes Ave., Pittsburgh, PA~15232.
%\label{CMU}}
%\addtocounter{address}{1}
%\altaffiltext{\theaddress}{Astrophysikalisches Institut Potsdam, Germany.
%\label{Potsdam}}
%%\addtocounter{address}{1}
%%\altaffiltext{\theaddress}{Department of Physics, Stanford University,
%   Stanford, CA 94305.
%\label{Stanford}}
% The abstract environment prints out the receipt and acceptance dates
% if they are relevant for the journal style.  For the aasms style, they
% will print out as horizontal rules for the editorial staff to type
% on, so long as the author does not include \received and \accepted
% commands.  This should not be done, since \received and \accepted dates
% are not known to the author.

\vbox{
\begin{abstract}
Low dispersion optical spectra have been obtained with
the Hobby-Eberly Telescope of 22 very red objects 
found in early imaging data from
the Sloan Digital Sky Survey.  The objects are assigned spectral types 
on the 2MASS system (Kirkpatrick et al. 1999) and are found to range from
late~M to late~L.  The red- and near-infrared colors from SDSS and 2MASS
correlate closely with each other, and most of the colors are closely
related to spectral type in this range; the exception is the
\hbox{$(i^* - z^*)$} color, which
appears to be independent of
spectral type between about~M7 and~L4.  The spectra
suggest that this independence
is due to the disappearance of the TiO and VO absorption in
the~$i$ band for later spectral types;
to the presence of strong Na~I and K~I absorption in the 
$i$ band; and to the gradual disappearance of the 8400 \AA{} absorption of
TiO and FeH in the~$z$ band.  
%For later spectral types, the colors become
%increasingly red with spectral type, with scatter in the near-infrared colors
%at mid-L.

\end{abstract}
}

\keywords{stars:low-mass, brown dwarfs - surveys}

\section{Introduction}

In the last five years, deep, wide area sky surveys carried out at red and
near-infrared wavelengths have discovered (among many other things) a large
number of stellar objects much cooler than the coolest previously-known 
late M dwarfs (see Reid \& Hawley 2000).  
This subject has long been of major astronomical interest
because of its relevance to star formation, planet formation, 
stellar evolution and the total mass density of the Galaxy.  These surveys,
the Deep Near-Infrared Survey (DENIS - Delfosse et al. 1997) and Two-Micron
All Sky Survey (2MASS - Skrutskie  1999) in $JHK_s$, and, more
recently, the Sloan Digital Sky Survey (SDSS - York et al. 2000) at
optical wavelengths, have led to the discovery of large numbers of
field brown dwarfs and the definition of two new spectral classes,
types L (Becklin \& Zuckerman 1988; Ruiz et al. 1997; 
Mart\'{\i}n et al. 1999, Kirkpatrick et al. 1999, 2000),
and T (Kirkpatrick et al. 1999; Burgasser  2001).  Broadly speaking,
the M$\rightarrow$L transition is marked by the disappearance of the
strong TiO and VO bands from the red spectrum and the L$\rightarrow$T
transition by the appearance of methane absorption at 1.6$\rm \mu m$ and 
2.2$\rm \mu m$ (Oppenheimer et al. 1995; Mart\'{\i}n et al. 1999; 
Kirkpatrick et al. 1999; Leggett et al. 2000; Burgasser 2001;
Reid 2001; Kirkpatrick 2001; Geballe et al.~2001).

The definition and delineation of these two spectral classes, their 
analysis via spectral modeling and the extension of the stellar mass function
to substellar masses have been the focus of much recent observational
and theoretical work.  Even with these new sky surveys, the identification
of candidate objects is difficult because they are so faint, and their
confirmation by spectroscopic observations requires a significant
amount of observing time
on large telescopes.  The SDSS has proven to be very effective at finding
candidate objects throughout the late M, L and T classes.  Although the SDSS
is a visible-wavelength survey and these very cool objects emit but a tiny
fraction of their luminosity at visible wavelengths, the reddest SDSS
bands,~$i$ and~$z$ (see Fukugita et al.~1996; Gunn et al.~1998)
have enabled the identification (not without some
difficulty, see below) of numerous faint red objects, both brown dwarfs
and high-redshift quasars.  Spectra of candidate objects selected by color are 
observed either by the SDSS spectrographs or with other telescopes, at
both optical and near-infrared wavelengths.  Early results on~L and~T dwarfs
based on SDSS data are given by Fan et al.~(1999), Strauss et al.~(1999),
and Leggett et al.~(2000).

This paper discusses results from
one of the spectroscopic efforts, optical-wavelength
observations of 22 SDSS red stellar candidate 
objects, made with the Hobby-Eberly Telescope (HET - Ramsey et al. 1998).
These objects have spectral types from~M7 to~L7, a range of
interest because the disappearance of the TiO and VO bands in this spectral 
region is very likely to be due to the condensation of these molecules into
solids, producing dusty atmospheres (Tsuji et al.
1996).

The objects in this paper are all of the late-type stars/brown dwarfs
observed with the HET during its commissioning/early operation phases,
and do not comprise a complete sample.
The next section describes the selection of target objects from the SDSS
imaging, and the HET spectroscopy is described in Section 3. The 
spectral typing is described in Section 4, and the relationships 
among spectral type, effective temperature, and colors are described 
in Section 5.  The conclusions are given in Section 6.  Additional discussions
of some of the objects in this paper are given by Geballe et al.~(2001),
Leggett et al.~(2001b), and Hawley et al. (2001, in preparation).

\section{Sloan Digital Sky Survey Imaging and Candidate Selection}

The Sloan Digital Sky Survey
uses a CCD camera \hbox{(Gunn et al. 1998)} on a
dedicated 2.5-m telescope \hbox{(Siegmund et al. 2001, unpublished)}
at Apache Point Observatory,
New Mexico, to obtain images in five broad optical bands over
10,000~deg$^2$ of the high Galactic latitude sky centered approximately
on the North Galactic Pole.  The five filters (designated $u$, $g$,
$r$, $i$, and~$z$) cover the entire wavelength range of the 
atmospheric transparency/CCD
response \hbox{(Fukugita et al. 1996;} \hbox{Gunn et al.~1998).}
Photometric calibration is provided by simultaneous
observations with a 20-inch telescope at the same site (Hogg et al.~2001).
The
survey data processing software measures the properties of each detected object
in the imaging data in all five bands, and determines and applies both
astrometric and photometric
calibrations (\hbox{Pier et al. 2001, unpublished}; 
\hbox{Lupton et al. 2001}).
At the time of this writing (Summer~2001) substantially more
than~1000~sq.~deg.~of sky have been imaged with the SDSS,
although some of the data
do not meet the strict survey
requirements for image quality.
The limiting magnitudes for these data are about 22.7, 21.8,
and 20.3 (5$\sigma$, point source) in $r$, $i$ and $z$ respectively
(very few of the red objects discussed herein
are detected in the~$u$ or~$g$ bands).

The automated image processing pipelines (Lupton et al.~2001)
used to process the imaging data find objects and merge the
observations of an object in each of the five bands; a 
formal flux density at the object position in bands in which the object is 
not detected is also measured. 
The software provides a set of flags for each
detected and measured object which describe the processing of that 
object and indicate the presence of any problems in the data or in the
data reduction.  These flags were used to select objects with reliable
data.
Many faint late-type dwarfs (and high-redshift quasars)
will be $z$-band-only detections, but at these faint levels single
band detections are highly contaminated by artifacts, mostly cosmic rays 
and ghosts of bright objects.  The objects for this study
were selected according to the following criteria:
\hbox{(a) ($i^*-z^*) > +1.6$}
(see Fan et al.~2000)\footnote{
Throughout this paper, measured SDSS magnitudes
will be denoted by $r^*$, $i^*$, and $z^*$
because of the preliminary nature of the 
photometric calibration, while the bands themselves are denoted by $r$, $i$,
and $z$}; (b)~at least a 2$\sigma$ detection in the $i$ band
and/or a detection in the 2MASS catalogue (although the released 2MASS
catalogue does not at present provide complete overlap with the SDSS);
and (c)~uncontaminated by substandard data (bad CCD columns and cosmic rays)
or by complex blending with the image of a nearby object. 

Candidates were selected from six SDSS imaging
runs: four equatorial scans (94, 125 in September~1998 and 752, 756 in
March~1999; see Fan et al.~2000 for details) covering slightly
over~500~deg$^2$, and two scans taken in October~1999 (SDSS imaging runs~1035
and~1043) that covered a combined area of~67~deg$^2$.  A total of~71 objects
was selected for the HET L~dwarf spectroscopy program; in 1999 and~2000
we obtained spectra of~22 of the candidates.  Since the HET operates in a
queue mode, the objects were selected for observation based solely on their
celestial position, so the~22 objects should provide a representative
subsample.

The~22 objects for which spectroscopic observations are reported herein are
listed in Table~1.  We give the object name, the spectral type as determined
in this paper, the $r^*$, $i^*$ and $z^*$ point spread function
magnitudes with their~1$\sigma$ errors and, where available, the 
2MASS $JHK_s$ magnitudes and their errors. Note that the SDSS and 2MASS
magnitudes are in different systems; the SDSS magnitudes are measured in the
$AB_{\nu}$ system (cf. Fukugita et al. 1996) whereas the 2MASS magnitudes
are in the Vega system.  In addition, the definition of SDSS magnitudes
is modified to measure magnitudes which are calculated from flux 
densities measured at low signal-to-noise ratios, or which
are formally zero or 
negative (Lupton, Gunn and Szalay 1999). The
calibration of the magnitudes is accurate to~0.03
in the~$r$ and~$i$ filters
and~0.05 in the~$z$ band (Stoughton et al.~2002).
The photometry has not
been corrected for foreground reddening (assumed to be negligible in the
three bands for these nearby objects).
The object name format is
\hbox{SDSSp Jhhmmss.ss$\pm$ddmmss.s}, where the coordinate
equinox is~J2000, and the ``p" refers to the preliminary nature of the
astrometry. The
estimated astrometric accuracies in each coordinate are~0.10$''$ rms. 
Object names will frequently
be abbreviated \hbox{as SDSShhmm+ddmm} in this paper. 
Figure~1 displays finding charts for the~22 objects made from the SDSS~$i$
images.  Note that these objects are likely to be nearby and to have 
large proper motions.  The last column of Table~1 contains the SDSS imaging run
number that is the basis for each set of coordinates;
the UT dates of the run numbers
are given in a footnote to the table.

Figure~2 shows the ($z^*$,$i^*-z^*$) color-magnitude diagram for the objects,
which are differentiated into three broad classes: (a)~earlier than
M9.5; (b)~L0-L3; and (c)~L4 and later (see Section 4).
Note that every observed candidate is indeed a late-type star/L~dwarf; 
there appears to be no contamination of our original sample by
other classes of sources.
The comparison sample consists of~50,000 high-latitude point sources 
randomly selected from a sample which contains objects which are
well-detected and classified by the image processing software as stars (point
sources) in the three most sensitive SDSS bands: $g$, $r$ and~$i$.
The ``blue'' ($i^*-z^* \sim$ 0) stars are halo and disk main sequence turn-off
stars, the red ($i^*-z^* \sim$ 1) stars are the disk M dwarfs.
The late~M and early~L dwarfs have $i^*-z^*$ = 1.6
to 2; the later L dwarfs have colors that are about one magnitude redder.

Figure~3 shows the $i^*-z^*$ vs. $r^*-i^*$ diagram for~18 of the objects;
the four sources not detected at the 3$\sigma$ level in the $r$ band
are not plotted.  The
comparison sample is 15,000 stellar objects matched with 2MASS (from
Finlator et al. 2000) which are relatively bright stars (because of the
different depths of the SDSS and 2MASS), and biased towards the redder
stars.  The late M and
L dwarfs are much redder in $i^*-z^*$ but tend to be bluer in $r^*-i^*$
than the values expected from the extrapolation of the colors
of warmer stars (cf. Fan et al.~2000).  

\section{Spectroscopy of L Dwarf Candidates}

Spectra of~22 SDSS L dwarf candidates were obtained
with the HET's Marcario
Low Resolution Spectrograph (LRS; Hill~et~al.~1998a,b;
Cobos~Duenas~et~al.~1998; Schneider~et~al.~2000) between May~1999
and December~2000.
The LRS is mounted in the Prime Focus Instrument Package, which
rides on the HET tracker.
The dispersive
element was \hbox{a 300 line mm$^{-1}$} grism blazed \hbox{at 5500 \AA .}
An OG515 blocking filter was installed to permit calibration of the spectra
beyond~8000~\AA .
The detector is a thinned, antireflection-coated
3072~$\times$~1024 Ford Aerospace CCD, and was
\hbox{binned $ 2 \times 2$} during readout; this produced
an image scale of~0.50$''$~pixel$^{-1}$ and a dispersion
\hbox{of $\approx$ 4.5 \AA\ pixel$^{-1}$}.  The spectra cover the
range from 5100--10,200~\AA\ at a resolution of approximately~20~\AA.

The wavelength calibration was provided by Ne, Cd, and~Ar comparison lamps; a
5$^{\rm th}$ order polynomial fit to the lines produced an rms error of less
than~1~\AA .
The relative flux calibration and atmospheric absorption band corrections
were performed by
observations of spectrophotometric standards, usually taken from
the primary
spectrophotometric standard list of Oke \& Gunn~(1983). 
The objects were observed under a wide range of conditions; the FWHM of the
spectra ranged from slightly under~2$''$ to over~3$''$.  The exposure times
varied from 347~s to 4380~s, with a median of 1400~s.
Absolute spectrophotometric calibration was performed
by scaling each spectrum so that
the~$i^*$ magnitudes synthesized from the spectra matched the SDSS photometric
measurements; this scaling used the SDSS response curves presented
by Fan et al.~(2001).  The spectra, rebinned to a linear wavelength scale
at \hbox{8~\AA\ pixel$^{-1}$}, are displayed in Figure~4.  Spectra of
four of the objects 
(SDSS0301+0020, SDSS0330+0000, SDSS0423$-$0414, and SDSS2555$-$0034)
are of markedly lower quality than the other 18 observations: the
low S/N spectra are binned at \hbox{16~\AA\ pixel$^{-1}$} in the figure.
The spectra are 
ordered by type, as discussed in the next section.
Only one of the~22 objects, SDSS0224$-$0721~(M8.5),
shows evidence for H$\alpha$ emission; this result is not surprising given
the low resolution (20~\AA ) and limited signal-to-noise ratio of the spectra
below 7000~\AA .

\section{Spectral Classification}

As part of the SDSS low-mass stars and brown dwarfs 
identification effort, we have developed both spectral
diagnostic and template fitting routines for M and L dwarfs;
this work is described in detail in our upcoming paper presenting
observations of several dozen late M and~L dwarfs
found by the SDSS (Hawley et al.~2001, in preparation).
The spectral diagnostics are taken from the Kirkpatrick et al.~(1999) 
prescription, augmented by several additional spectral features which 
proved more robust for lower spectral resolution, and often lower 
signal-to-noise ratio, spectra.  The spectral types 
were assigned through a combination of the spectral diagnostic and 
template fitting results; the spectral types were
independently checked by two of us (SLH, KRC)
by visually comparing each HET
spectrum with the Keck standards from Kirkpatrick et al. (1999).  
We estimate the uncertainty in the spectral types for the 18 stars 
with reasonable signal-to-noise ratio
spectra to be about $\pm 1$ spectral type, 
while the spectral types of the four stars with poor signal-to-noise spectra
are only suggestive and should be regarded as quite uncertain; these
spectral types are denoted by a colon in
Table~1.  The objects presented in this paper
have spectral classes ranging from~M7 to~L7.

The HET observation of one of the objects, SDSS1430+0013, was presented by 
Schneider et al. (2000); in that paper a preliminary classification
of~L0 was assigned, while this paper classifies the object as~M8.
Observations of a second object, SDSS0413$-$0114, were
presented by Fan et al. (2000); the visual classification of that spectrum
was L0, compared to L0.5 in Table 1 of this paper.
SDSS0205+1251 is a 2MASS~L dwarf, assigned 
spectral type L5 by Kirkpatrick et al. (2000); we classify it as L4
based on the HET data.
In addition, four of the objects discussed in the present paper are
described by Geballe et al. (2001):
SDSS0107+0041
(L7 in the present paper) is classified as~L5.5 by Geballe et al. (2001);
SDSS0236+0048~(L6) as~L6.5; SDSS0423$-$0414~(L5:) as~T0; and
SDSS2255$-$0034 (L0:) as M8.5.  The only
discrepancy of note is for SDSS0423$-$0414,
but the spectrum presented in this
paper (see Figure~4) is of low signal-to-noise ratio, and
hence the classification in Table~1 is only suggestive as discussed above.
In general our results are in concurrence with the
results of
Reid et al.~(2001), Testi et al.~(2001) and Geballe et al.~(2001) which show 
that the M$\rightarrow$L spectral sequence can be consistently
defined by both optical and infrared spectroscopy.

\section{Discussion}

The objects observed in this study were selected to have very red
\hbox{($i^*-z^* > +1.6$)} colors.  
%As Figure~3 shows, the $r^*-i^*$ color saturates or turns
%over at about \hbox{$r^* - i^* \approx 2 - 2.5$}
%for objects of spectral type very late~M and later.  This
%happens because the red VO and TiO bands, instead of becoming progressively
%deeper with decreasing temperature, become shallower.
%The depth of the 
%TiO$\lambda$7053 feature peaks at about M8-M9, then declines in 
%strength and disappears by about L2-L3 (Jones \& Tsuji 1997; Kirkpatrick et
%al. 1999); this is attributed to the precipitation of these species into
%dust (Tsuji et al. 1996; Burrows, Marley \& Sharp  2000).
In the top panel of Figure~5,
we plot the SDSS $i^*-z^*$ color versus spectral type, including
the data both from the present paper and from Fan et al.~(2000).
The $i^*-z^*$ color is roughly constant at~+1.8 (albeit with large scatter)
until spectral type L2-L3, then steadily
increases to nearly~+3 for late L dwarfs.  
While the \hbox{$i^* - z^*$} color is not good indication of spectral
class for objects of~M8 through~L3, colors based on $z^*$ and one of the
infrared bands can be used for spectral classification for all types
from late~M to late~L.  The bottom panel in Figure~5 shows the relationship
between spectral type and \hbox{$z^* - J$} (top); a plot of
\hbox{$z^* - K_s$} vs. spectral type has a similar appearance.

What might produce the apparent lack of correlation between
\hbox{$i^* - z^*$} color and spectral type in early~L dwarfs?
One possibility is that this result is merely an
artifact of our selection technique.  Since the sample selection required
that \hbox{$i^* - z^* > +1.6$,} any objects bluer than that limit cannot
appear in Figure~5.
We consider this explanation unlikely, however,
given the distribution of the observations; the \hbox{$i^* - z^*$} measurements
of the early~L dwarfs cluster around a value of~+1.9, with very few points
bluer than~+1.7. In any case, \hbox{$i^* - z^*$} cannot be an effective
spectral type indicator in this range, for if our selection has missed
a ``blue" population, then the \hbox{$i^*-z^*$} variation is quite
substantial.
Figure~5 shows that the
$i^*-z^*$ color is, at least as far as the present data are concerned,
only a very crude predictor of spectral type;
it allows identification of cool substellar objects but not the measurement
of spectral type.  

Figure~6 displays the spectra of SDSS0411$-$0556
(M8.5) and SDSS0107+0041 (L7) compared with the
filter responses of the SDSS $r$, $i$, and $z$ filters.  
For late~L and 
T dwarfs, the flux shortwards of about 8000~\AA\ is almost completely
suppressed by the extremely broad wings of the NaD $\lambda$5889/5896
and K~I $\lambda$7665/7699 resonance lines
(Tsuji et al. 1996; Kirkpatrick et al. 1999; Liebert et al. 2000;
Burrows et al. 2000), as can be seen for spectra of type L4 and greater
in Figure~4.  
The $i^*-z^*$ color of these
late-type dwarfs is strongly influenced by two features: the removal
of flux from the $i$ band for objects of about L2 and later by Na~I and 
K~I absorption: and strong absorption due to the TiO band at 8432 \AA{}
and FeH at 8692 \AA{}.  As Figure~4 and Kirkpatrick et al. (1999) show, 
these absorption bands peak at spectral types of about L0 and L3 
respectively, and disappear at spectral types of about L2 and L5 
respectively.  An understanding of how these absorption features interplay with 
the SDSS filters as a function of temperature is
complicated by the fact that because the bands occur close to the peak
sensitivity of the SDSS $z$ filter, the differences in the detailed
$z$ response of the individual columns of the SDSS camera and changes in
the atmospheric water vapor column density during the observations
will introduce scatter in the observations (and may account for some
fraction of the variations seen in Figure~5). 
This complex issue will be further investigated in
future work with a much larger sample of SDSS brown dwarfs
(Hawley et al. 2001 in preparation).

Figure~7 shows optical-near infrared color-color diagrams for the objects
with published 2MASS magnitudes from Fan et al. (2000) and the present paper.
Both the $H - K_s$ and $J - K_s$ colors versus $i^*-z^*$ show, with the
exception of one object in $J-K_s$, a good correlation between the
infrared and optical colors and a steady progression towards redder colors
at both optical and infrared wavelengths.  
The outlying object, SDSS/2MASS0205+1251, is much redder in $J-K_s$
relative to $i^*-z^*$ than expected (Figure~7a). 
Since the $H-K_s$ color of this object is close to the
expected value (Figure~7b), this implies a significant reduction in $J$
flux.  We are not aware of a published near-infrared spectrum of this object;
we speculate
that its 1.15$\mu$m H$_2$O feature may be deeper than for objects of
neighboring spectral type.  The spectra of dwarfs of late M to
early L spectral type
are difficult to fit without including the effects of photospheric
dust extinction, which makes the strong water bands shallower than 
expected from dust-free model atmospheres, whereas fits to late L and
T dwarfs show that dust does not affect the model colors very much
(Allard~1998; Jones \& Tsuji~1997; Leggett et al.~2001a;
Marley \& Ackerman~2001).  Thus the region around L4-L5
may be where dust is causing some
scatter in the infrared colors.

\section{Conclusions}

We have obtained low-dispersion red spectra with the Hobby-Eberly telescope 
of~22 very red point source objects selected from early SDSS imaging data.
Spectral types on the Kirkpatrick
et al. (1999, 2000) system can be assigned with reasonable confidence to
18 of the objects, with the spectra for the other four having poor signal-to-
noise ratio.  We find:

\begin{itemize}

\item
The 22 objects include nine late type M dwarfs (M7.5 - M9.5) and 13 L dwarfs, 
ranging from spectral types L0 to~L7.  The spectral types are in good
agreement with those assigned in other work.

\item
The spectra show strong suppression of the optical-wavelength flux due 
to absorption in the wings of the Na~I and K~I lines (cf. Kirkpatrick et al.
1999, 2000; Burrows, Marley \& Sharp 2000; and Liebert et al. 2000) for 
spectral types later than about~L3.

\item
The SDSS ($i^*-z^*$) colors from this paper and from Fan et al.~(2000)
show no trend with spectral types between~M8 and~L3.
It is possible, but unlikely,
that this is a result of our selection procedure.
The increasing red color for spectral types~L4 and later can be
attributed to the suppression of the~$i$ flux by Na~I and K~I
absorption and the disappearance of the strong TiO and FeH absorption from the
$z$ flux.

\item
The near-infrared colors ($J-K_s$ and $H-K_s$) from 2MASS are closely
correlated with the $i^*-z^*$ colors.
%It is possible that atmospheric models which reproduce the $i^*z^*JHK_s$
%colors for these objects will suggest a re-assignment or re-definition
%of spectral types.

\item
The present paper contains only a small number of objects of type L4 and
later.  One object, SDSS/2MASS0205+1251
(L4) seems to have a weaker than expected
$J$-band flux.  Since the depth of the water bands is anticorrelated with
the amount of dust present in the photosphere, the anomalous $J$-band
measurement may arise from
the absence of atmospheric dust in this object.

\end{itemize}

\acknowledgments

We would like to thank Hugh Harris for several useful comments on this
study.
This work was supported in part by National Science Foundation grants
AST99-00703~(DPS and~GTR), PHY00-70928~(XF), and AST00-71091~(MAS).
XF acknowledges
support from a Frank and Peggy Taplin Fellowship. GK thanks Princeton
University and NASA (NAG-6734) for support.

The Sloan Digital Sky Survey
\footnote{The SDSS Web site \hbox{is {\tt http://www.sdss.org/}.}}
(SDSS) is a joint project of The University
of Chicago, Fermilab, the Institute for Advanced Study, the Japan
Participation Group, The Johns Hopkins University, the
Max-Planck-Institute for Astronomy (MPIA), the Max-Planck-Institute for
Astrophysics
(MPA), New Mexico State University, Princeton University, the United
States Naval Observatory, and the University of Washington. Apache
Point Observatory, site of the SDSS telescopes, is operated by the
Astrophysical Research Consortium (ARC). 
Funding for the project has been provided by the Alfred~P.~Sloan
Foundation, the SDSS member institutions, the National Aeronautics and
Space
Administration, the National Science Foundation, the U.S.~Department of
Energy, the Japanese Monbukagakusho, and the Max Planck Society.

The Hobby-Eberly Telescope (HET) is a joint project of the University of Texas
at Austin,
the Pennsylvania State University,  Stanford University,
Ludwig-Maximillians-Universit\"at M\"unchen, and Georg-August-Universit\"at
G\"ottingen.  The HET is named in honor of its principal benefactors,
William P. Hobby and Robert E. Eberly.  The Marcario Low-Resolution
Spectrograph is named for Mike Marcario of High Lonesome Optics, who
fabricated several optics for the instrument but died before its completion;
it is a joint project of the Hobby-Eberly Telescope partnership and the
Instituto de Astronom\'{\i}a de la Universidad Nacional Aut\'onoma de M\'exico.

This publication makes use of data products from the Two Micron All Sky Survey,
which is a joint project of the University of Massachusetts and the Infrared
Processing and Analysis Center/California Institute of Technology,
funded by the National Aeronautics and Space Administration and the 
National Science Foundation.

\newpage
\centerline{\bf Figure Captions}

%\begin{figure}
%\plotfiddle{qpairf1.ps}{6.5in}{0.0}{90.0}{90.0}{-255.0}{-32.0}
\figcaption{
Finding charts for the~22 objects discussed in this paper;
each individual
chart is~100$''$ on a side.  All frames are~$i$ images taken with the SDSS
camera.  The small arrow in the lower left of each chart indicates the
direction of north; all charts have ``sky" parity, so east is
located~90$^{\circ}$ counterclockwise from north.
\label{fig1}}
%\end{figure}

\figcaption{
Color-magnitude plot ($z^*$ vs. $i^*-z^*$)
for the candidate L dwarfs, compared with data for 50,000 anonymous 
high-latitude stars (extracted from the SDSS data to be well detected and with
point-source morphology in the SDSS $g$, $r$ and $i$ filters. Open stars:
M dwarfs.  Open circles: L0-L3.  Filled circles: L4 and later.
\label{fig2}}

\figcaption{
SDSS Color-color plot for the 18
objects detected in all three of the $r$, $i$ and $z$ bands,
compared with SDSS data for 15,000 stars from Finlator et al.~(2000).  The
symbols are the same as in Figure~2.  
\label{fig3}}

%\begin{figure}
%\plotfiddle{qpairf3.ps}{5.0in}{90.0}{80.0}{80.0}{345.0}{-15.0}
\figcaption{
HET LRS spectra of the 22 dwarfs ordered by spectral type.  The spectral
resolution is approximately 20~\AA\ and the data have been rebinned
\hbox{to 8 \AA\ pixel$^{-1}$.}  The four spectra with low signal-to-noise
ratio
(SDSS0301+0020, SDSS0330+0000, SDSS0423$-$0414, and SDSS2555$-$0034)
are binned \hbox{at 16 \AA\ pixel$^{-1}$.}
%\hbox{$2.29 \times 10^{-31}$ erg cm$^{-2}$ s$^{-1}$ Hz$^{-1}$.}
\label{fig4}}
%\end{figure}

\figcaption{Top panel:
SDSS ($i^*-z^*$) color vs. spectral type; the circles represent
data from Table 1 and the crosses are objects from
Fan et al.~(2000).  The open circles are the four objects
with uncertain spectral type. The points corresponding to objects assigned 
the same spectral type have been slightly offset horizontally for clarity.
Bottom panel: \hbox{$z^* - J$} vs. spectral types; the symbols are the
same as in the top panel.  Note that $z^*$ is an AB based system, while~$J$
measurements are normalized to Vega.
\label{fig5}}

\figcaption{
SDSS $r$, $i$, and $z$ relative system responses (including~1.3 airmasses)
compared with
HET spectra of SDSS0411$-$0556~(M8.5) and SDSS0107+0041~(L7).
The strong absorption
between about 8500~\AA\ and 8800~\AA\ seen in SDSS0411$-$0556 is due to the
8432~\AA\ and  8692~\AA\ TiO and FeH bands, which disappear at about L3-L4
(see Figure~4).
\label{fig6}}

\figcaption{
(Upper Panel) ($J-K_s$) vs. ($i^*-z^*$)  and (lower panel)
($H-K_s$) vs. ($i^*-z^*$) for objects from
the present paper and Fan et al. (2000) with detections in the released 2MASS
data.  Crosses: type M.  Open symbols: L0-L3.  Filled symbols:
L4 and later.
\label{fig7}}

\clearpage

\begin{figure}
\plotfiddle{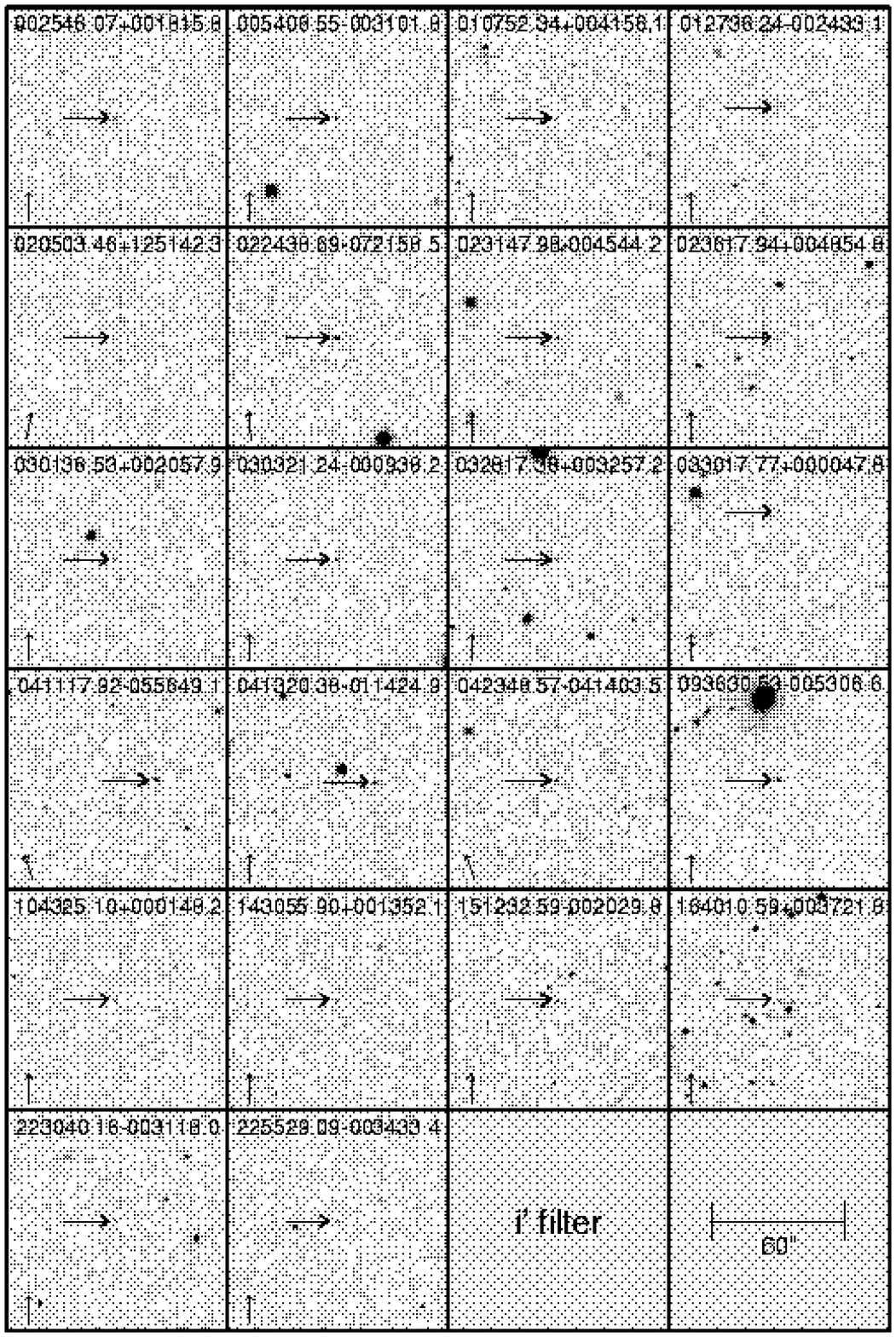}{8.0in}{0.0}{95.0}{95.0}{-240.0}{-80.0}
%\label{ }
\end{figure}
\clearpage

\begin{figure}
\plotfiddle{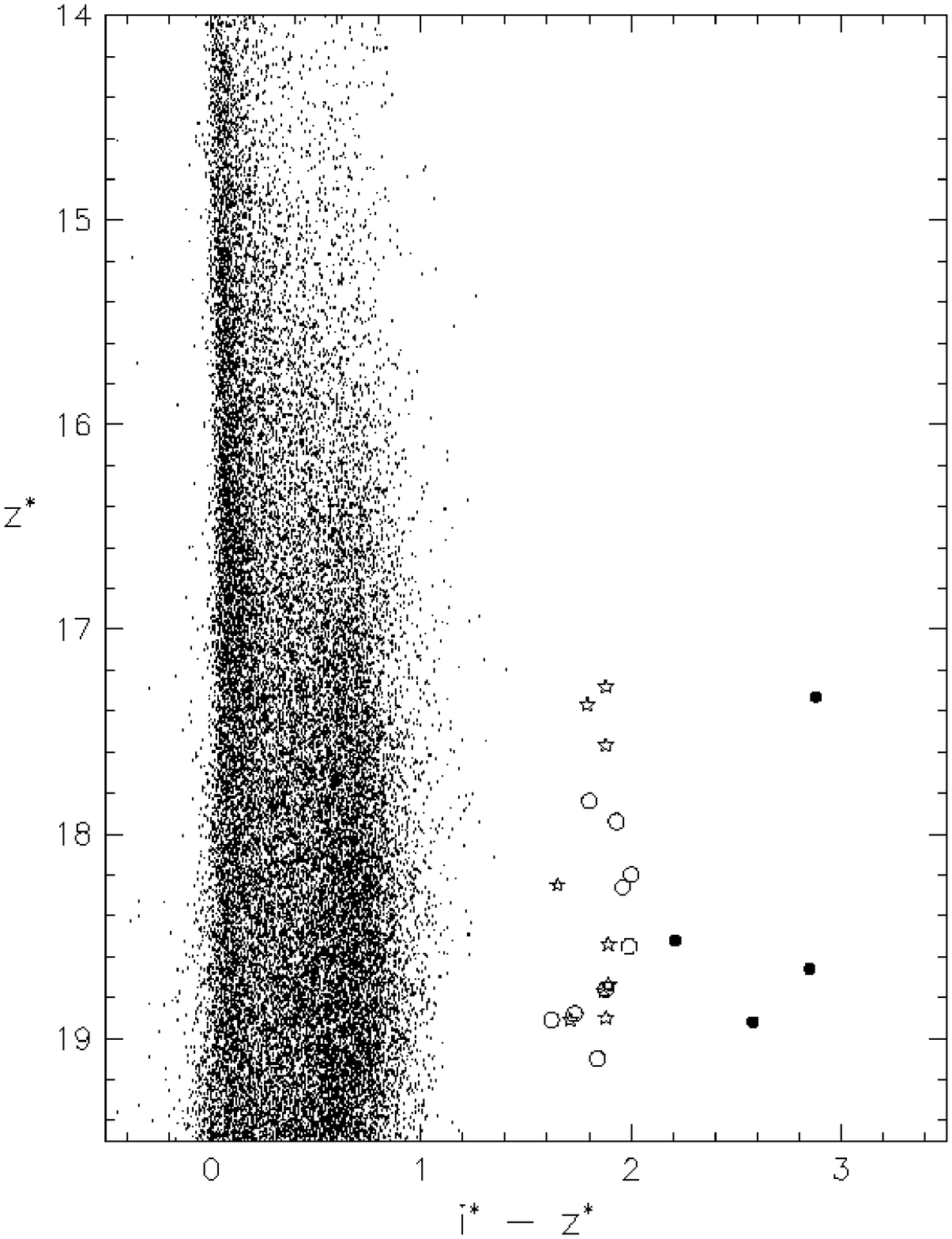}{8.0in}{0.0}{80.0}{80.0}{-250.0}{0.0}
%\label{ }
\end{figure}
\clearpage

\begin{figure}
\plotfiddle{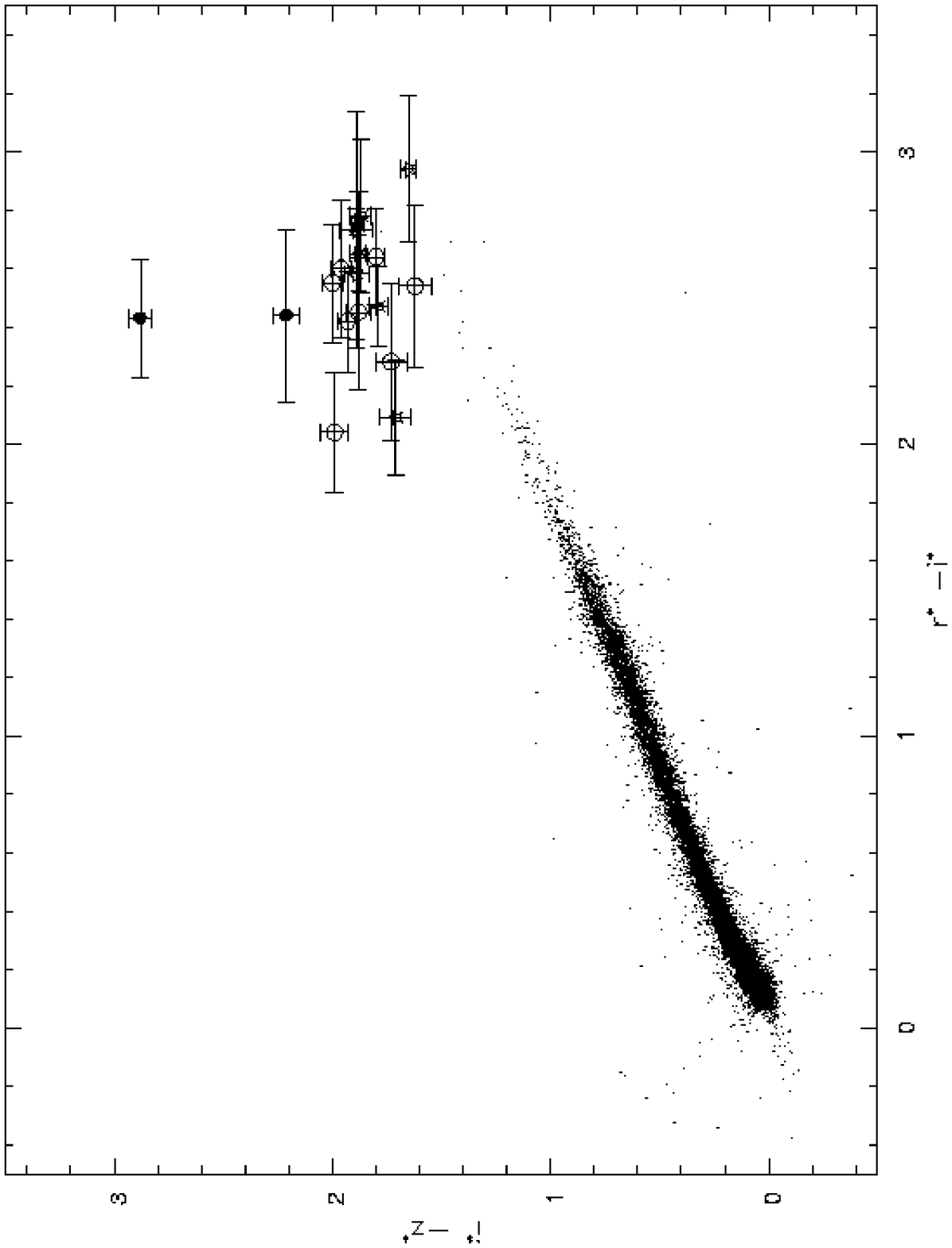}{8.0in}{0.0}{80.0}{80.0}{-250.0}{0.0}
%\label{ }
\end{figure}
\clearpage

\begin{figure}
\plotfiddle{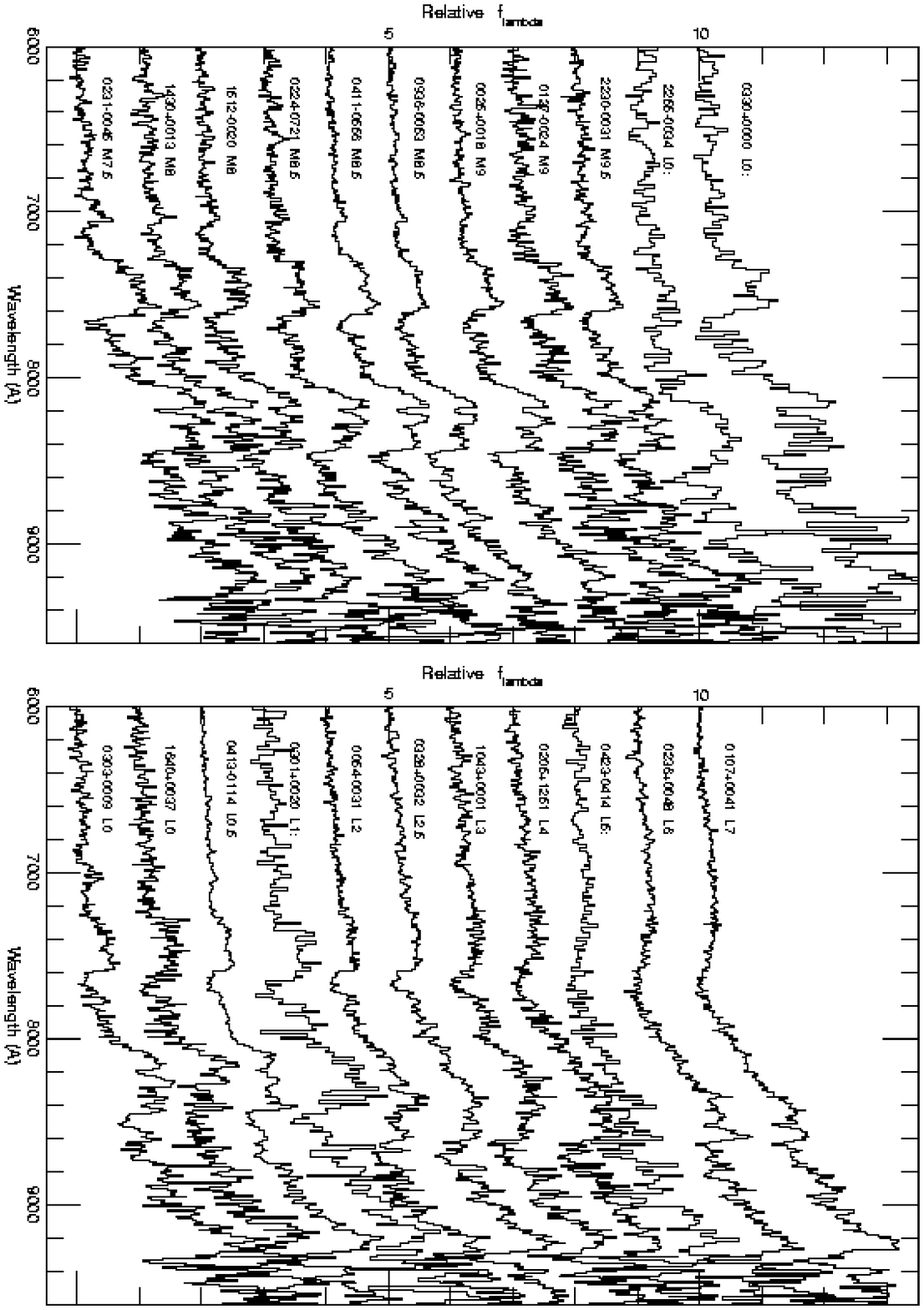}{8.0in}{180.0}{80.0}{80.0}{290.0}{600.0}
%\label{ }
\end{figure}
\clearpage

\begin{figure}
\plotfiddle{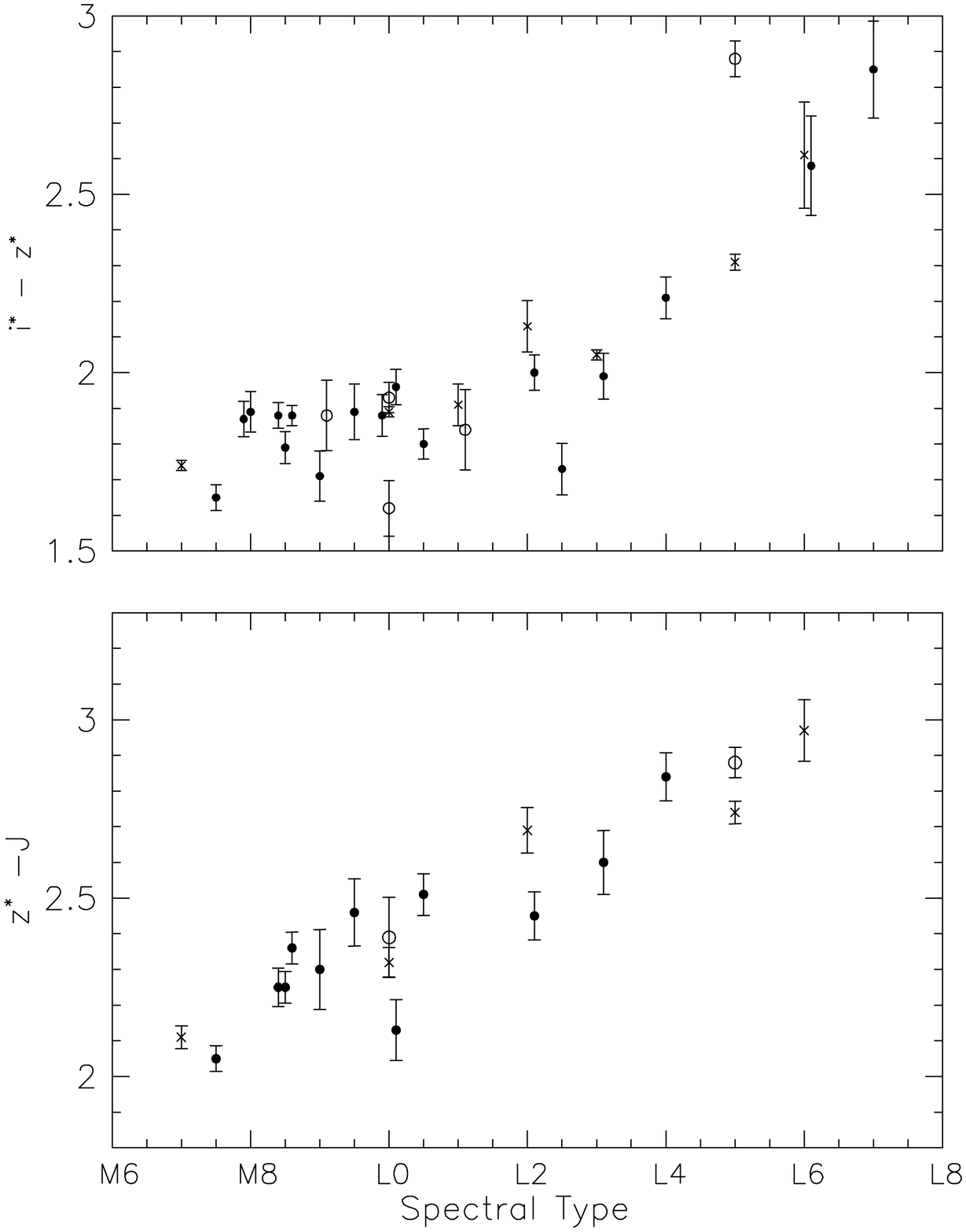}{8.0in}{0.0}{80.0}{80.0}{-250.0}{0.0}
%\label{ }
\end{figure}
\clearpage

\begin{figure}
\plotfiddle{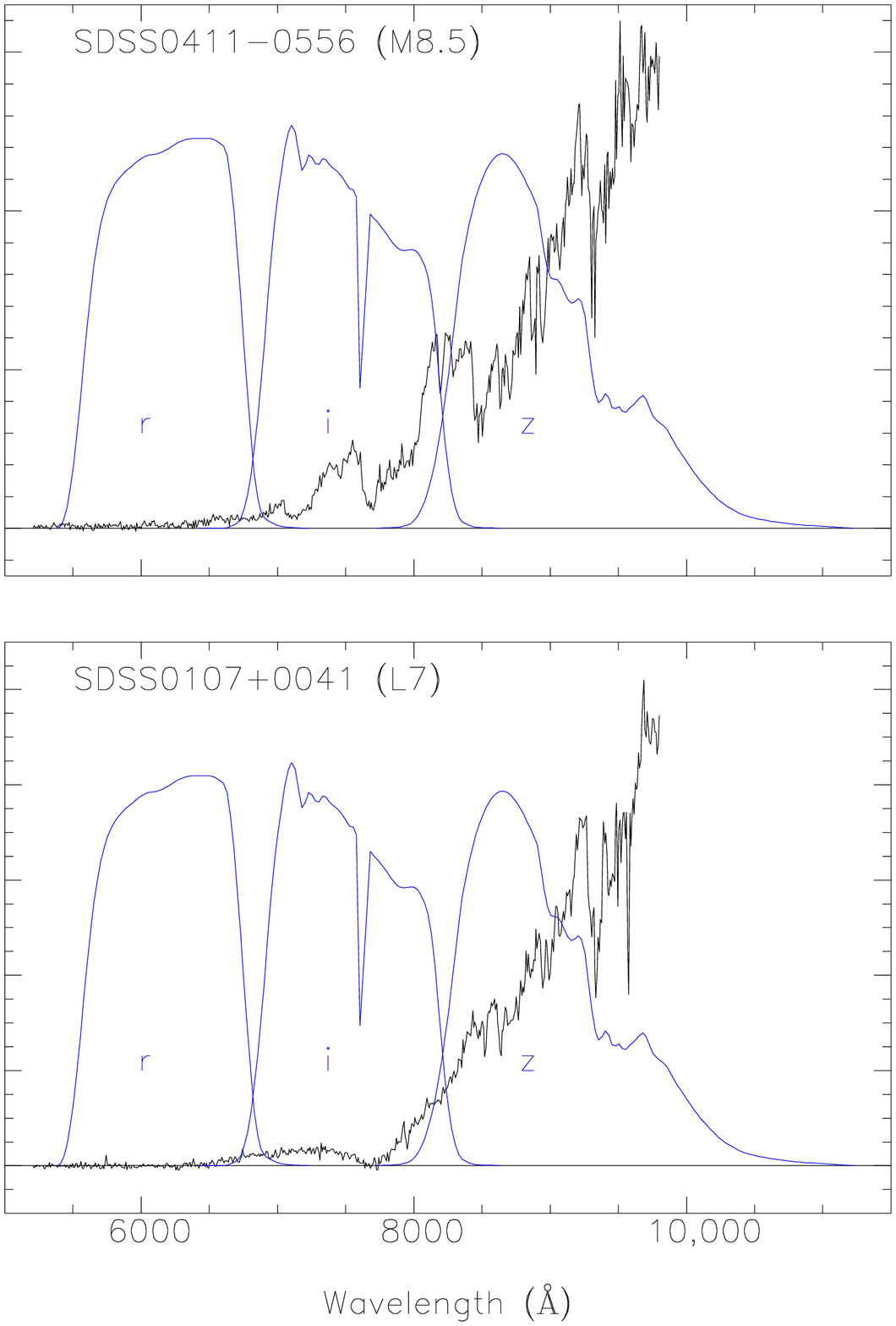}{8.0in}{0.0}{80.0}{80.0}{-250.0}{0.0}
%\label{ }
\end{figure}
\clearpage

\begin{figure}
\plotfiddle{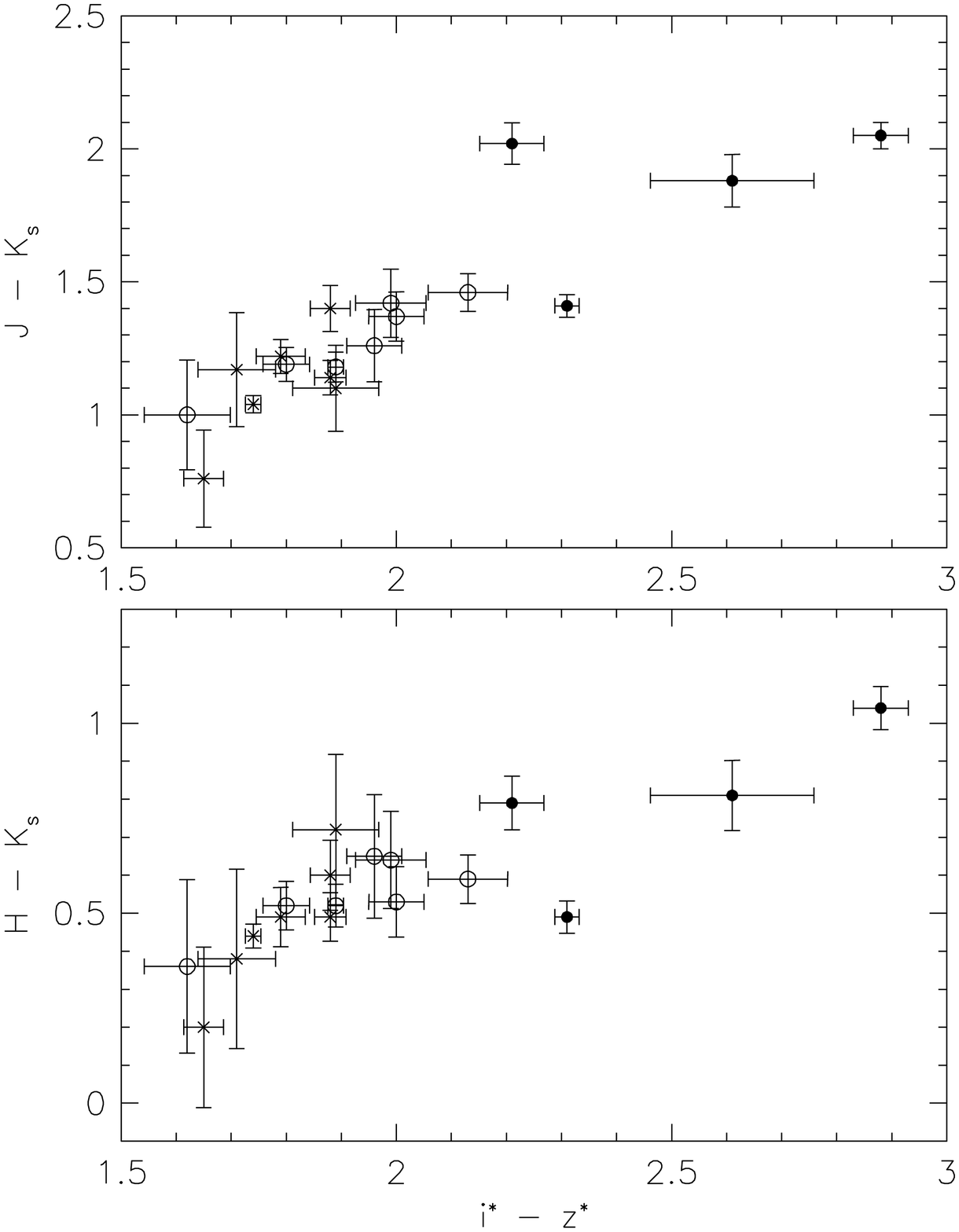}{8.0in}{0.0}{80.0}{80.0}{-250.0}{0.0}
%\label{ }
\end{figure}
\clearpage

%-------- Insert Table 1 below this line -----------

\begin{scriptsize}

\halign{\hskip 12pt
# \hfil \tabskip=0em plus0em minus0em
& # \hfil
&\hfil # \hfil 
&\hfil # \hfil 
&\hfil # \hfil 
&\hfil # \hfil 
&\hfil # \hfil 
&\hfil # \hfil 
& \hfil #  \cr
\multispan{9}{\hfil TABLE 1. Positions and Photometry of SDSS
 L Dwarfs \hfil}\cr
\noalign{\bigskip\hrule\smallskip\hrule\medskip}
\hfil Object \hfil & \hfil Spec \hfil
&&&&&&& \hfil SDSS \hfil \cr
\hfil (SDSSp) \hfil & \hfil Type \hfil
&\hfil $r^*$ \hfil & \hfil $i^*$ \hfil
&\hfil $z^*$ \hfil & \hfil $J$ \hfil & \hfil $H$ \hfil 
&\hfil $K_s$ \hfil & \hfil Run \hfil \cr
\noalign{\medskip\hrule\bigskip}
J002546.07$+$001815.8 & M9  
 & 24.71 $\pm$ 0.60
 & 20.78 $\pm$ 0.07
 & 18.90 $\pm$ 0.07
& --- & --- & --- &   94 \cr
J005406.55$-$003101.8 & L2  
 & 22.75 $\pm$ 0.20
 & 20.20 $\pm$ 0.04
 & 18.20 $\pm$ 0.03
 & 15.75 $\pm$ 0.06
 & 14.91 $\pm$ 0.06
 & 14.38 $\pm$ 0.07
&   94 \cr
J010752.34$+$004156.1 & L7  
 & 23.98 $\pm$ 0.59
 & 21.51 $\pm$ 0.13
 & 18.66 $\pm$ 0.04
& --- & --- & --- &   94 \cr
J012736.24$-$002433.1 & M9  
 & 22.71 $\pm$ 0.19
 & 20.62 $\pm$ 0.05
 & 18.91 $\pm$ 0.05
 & 16.61 $\pm$ 0.10
 & 15.82 $\pm$ 0.14
 & 15.44 $\pm$ 0.19
&   94 \cr
J020503.46$+$125142.3 & L4  
 & 23.17 $\pm$ 0.29
 & 20.73 $\pm$ 0.05
 & 18.52 $\pm$ 0.03
 & 15.68 $\pm$ 0.06
 & 14.45 $\pm$ 0.05
 & 13.66 $\pm$ 0.05
& 1035 \cr
\noalign{\smallskip}
J022438.69$-$072158.5 & M8.5
 & 21.63 $\pm$ 0.13
 & 19.16 $\pm$ 0.04
 & 17.37 $\pm$ 0.02
 & 15.12 $\pm$ 0.04
 & 14.39 $\pm$ 0.06
 & 13.90 $\pm$ 0.05
& 1045 \cr
J023147.98$-$004544.2 & M7.5
 & 22.84 $\pm$ 0.25
 & 19.90 $\pm$ 0.03
 & 18.25 $\pm$ 0.02
 & 16.20 $\pm$ 0.03
 & 15.64 $\pm$ 0.11
 & 15.44 $\pm$ 0.18
&  125 \cr
J023617.94$+$004854.8 & L6  
 & 24.83 $\pm$ 0.61
 & 21.50 $\pm$ 0.13
 & 18.92 $\pm$ 0.05
& --- & --- & --- &   94 \cr
J030136.53$+$002057.9 & L1: 
 & 23.79 $\pm$ 0.41
 & 20.94 $\pm$ 0.08
 & 19.10 $\pm$ 0.08
& --- & --- & --- &   94 \cr
J030321.24$-$000938.2 & L0  
 & 22.82 $\pm$ 0.23
 & 20.22 $\pm$ 0.04
 & 18.26 $\pm$ 0.03
 & 16.13 $\pm$ 0.08
 & 15.52 $\pm$ 0.12
 & 14.87 $\pm$ 0.11
&   94 \cr
\noalign{\smallskip}
J032817.38$+$003257.2 & L2.5
 & 22.89 $\pm$ 0.26
 & 20.61 $\pm$ 0.06
 & 18.88 $\pm$ 0.04
& --- & --- & --- &  125 \cr
J033017.77$+$000047.8 & L0: 
 & 23.07 $\pm$ 0.27
 & 20.53 $\pm$ 0.06
 & 18.91 $\pm$ 0.05
 & 16.52 $\pm$ 0.10
 & 15.88 $\pm$ 0.14
 & 15.52 $\pm$ 0.18
&   94 \cr
J041117.92$-$055649.1 & M8.5
 & 21.91 $\pm$ 0.11
 & 19.16 $\pm$ 0.02
 & 17.28 $\pm$ 0.02
 & 14.92 $\pm$ 0.04
 & 14.27 $\pm$ 0.04
 & 13.78 $\pm$ 0.05
& 1045 \cr
J041320.38$-$011424.9 & L0.5
 & 22.28 $\pm$ 0.16
 & 19.64 $\pm$ 0.03
 & 17.84 $\pm$ 0.03
 & 15.33 $\pm$ 0.05
 & 14.66 $\pm$ 0.05
 & 14.14 $\pm$ 0.06
&  125 \cr
J042348.57$-$041403.5 & L5: 
 & 22.64 $\pm$ 0.20
 & 20.21 $\pm$ 0.04
 & 17.33 $\pm$ 0.03
 & 14.45 $\pm$ 0.03
 & 13.44 $\pm$ 0.04
 & 12.40 $\pm$ 0.04
& 1045 \cr
\noalign{\smallskip}
J093630.53$-$005306.6 & M8.5
 & 22.10 $\pm$ 0.12
 & 19.45 $\pm$ 0.03
 & 17.57 $\pm$ 0.02
 & 15.32 $\pm$ 0.05
 & 14.52 $\pm$ 0.06
 & 13.92 $\pm$ 0.07
&  756 \cr
J104325.10$+$000148.2 & L3  
 & 22.58 $\pm$ 0.20
 & 20.54 $\pm$ 0.05
 & 18.55 $\pm$ 0.04
 & 15.95 $\pm$ 0.08
 & 15.17 $\pm$ 0.08
 & 14.53 $\pm$ 0.10
&  752 \cr
J143055.90$+$001352.1 & M8  
 & 23.01 $\pm$ 0.22
 & 20.43 $\pm$ 0.04
 & 18.54 $\pm$ 0.04
& --- & --- & --- &  756 \cr
J151232.59$-$002029.8 & M8  
 & 23.42 $\pm$ 0.26
 & 20.64 $\pm$ 0.04
 & 18.77 $\pm$ 0.03
& --- & --- & --- &  752 \cr
J164010.59$+$003721.8 & L0  
 & 23.09 $\pm$ 0.26
 & 20.64 $\pm$ 0.05
 & 18.76 $\pm$ 0.03
& --- & --- & --- &  752 \cr
\noalign{\smallskip}
J223040.16$-$003118.0 & M9.5
 & 23.36 $\pm$ 0.40
 & 20.63 $\pm$ 0.06
 & 18.74 $\pm$ 0.05
 & 16.28 $\pm$ 0.08
 & 15.90 $\pm$ 0.14
 & 15.18 $\pm$ 0.14
&   94 \cr
J225529.09$-$003433.4 & L0: 
 & 22.29 $\pm$ 0.17
 & 19.87 $\pm$ 0.03
 & 17.94 $\pm$ 0.03
& --- & --- & --- &   94 \cr
\noalign{\medskip\hrule}}

\medskip\noindent
Notes: Positions are in J2000.0 coordinates.
The SDSS photometry is reported in terms of asinh magnitudes;
see Lupton, Gunn, \& Szalay~(1999) for details.
In this system, zero flux
corresponds to 24.8, 24.4, and~22.8
in $r^*$, $i^*$, and~$z^*$, respectively.
Note that SDSS magnitudes are AB-based and the infrared
magnitude system is based on Vega.
Spectral types followed by a colon have estimated errors of
two subclasses.

\noindent
Dates of SDSS Imaging Runs:
\hbox{94 (19 Sept 1998),}
\hbox{125 (25 Sept 1998),}
\hbox{752 (21 Mar 1999),}
\hbox{756 (22 Mar 1999),}
\hbox{1035 (12 Oct 1999),} and
\hbox{1043 (13 Oct 1999)}
\end{scriptsize}

\end{document}